# Distributed Storage Codes with Repair-by-Transfer and Non-achievability of Interior Points on the Storage-Bandwidth Tradeoff

Nihar B. Shah, K. V. Rashmi, P. Vijay Kumar, *Fellow, IEEE*, and
Kannan Ramchandran, *Fellow, IEEE*

## Abstract

Regenerating codes are a class of recently developed codes for distributed storage that, like Reed-Solomon codes, permit data recovery from any subset of $k$ nodes within the $n$-node network. However, regenerating codes possess in addition, the ability to repair a failed node by connecting to an arbitrary subset of $d$ nodes. It has been shown that for the case of functional-repair, there is a tradeoff between the amount of data stored per node and the bandwidth required to repair a failed node. A special case of functional-repair is exact-repair where the replacement node is required to store data identical to that in the failed node. Exact-repair is of interest as it greatly simplifies system implementation.

The first result of the paper is an explicit, exact-repair code for the point on the storage-bandwidth tradeoff corresponding to the minimum possible repair bandwidth, for the case when $d = n - 1$. This code has a particularly simple graphical description and most interestingly, has the ability to carry out exact-repair through mere transfer of data and without any need to perform arithmetic operations. Hence the term *repair-by-transfer*.

The second result of this paper shows that the interior points on the storage-bandwidth tradeoff cannot be achieved under exact-repair, thus pointing to the existence of a separate tradeoff under exact-repair. Specifically, we identify a set of scenarios, termed 'helper node pooling', and show that it is the necessity to satisfy such scenarios that over-constrains the system.

Nihar B. Shah, K. V. Rashmi and P. Vijay Kumar are with the Dept. of ECE, Indian Institute Of Science, Bangalore, India. Email: {nihar, rashmikv, vijay}@ece.iisc.ernet.in. P. Vijay Kumar is also an adjunct faculty member of the Electrical Engineering Systems Department at the University of Southern California, Los Angeles, CA 90089-2565. Kannan Ramchandran is with the Dept. of EECS, University of California, Berkeley, USA. Email: kannanr@eecs.berkeley.edu.

# I. Introduction

In a distributed storage system, information pertaining to a data file is dispersed across nodes in a network in such a manner that an end-user (termed as the data-collector or DC) can retrieve the data stored by tapping into a subset of nodes in the network. A popular option that reduces network congestion and leads to increased resiliency in the face of node failures, is to employ erasure coding, for example by calling upon maximum-distance-separable (MDS) codes such as Reed-Solomon (RS) codes.

Let the data stored in the network be represented by a collection of $B$ message symbols, with each message symbol drawn from a finite field $\mathbb{F}_q$ of size $q$. With RS codes, data is stored across $n$ nodes in the network in such a way that the entire data can be recovered by a data-collector by connecting to any arbitrary $k = B$ nodes, a process of data recovery that we will refer to as *reconstruction*. Several distributed storage systems such as RAID-6 [1], OceanStore [2] and Total Recall [3] employ such an erasure-coding option.

Upon failure of an individual node, a self-sustaining data storage network must necessarily possess the ability to *regenerate* (i.e., repair) the failed node. An obvious means of accomplishing this task is by first permitting the replacement node to download the entire data stored in any $k$ nodes and then proceeding to extract the data that was stored in the failed node. Such a procedure is indeed mandated when RS codes are employed to distribute the data and nodes are restricted to carry out linear operations on the data stored within them.

RS codes treat the data stored in each node as a single symbol belonging to the finite field $\mathbb{F}_q$. When this is coupled with the restriction that individual nodes perform linear operations over $\mathbb{F}_q$, it follows that the smallest unit of data that can be downloaded from a node assisting in the repair of a failed node equals the amount of information stored in the node itself, i.e., is the equivalent of an $\mathbb{F}_q$ symbol. As a consequence of the MDS property of an RS code, when carrying out repair of a failed node, the replacement node must necessarily collect data from at least $k$ other nodes. As a result, it follows that the total amount of data download needed to repair a failed node can be no smaller than $B$, the size of the entire message.

However, downloading the entire message of size $B$ in order to repair a single node that stores only a fraction of the entire data is wasteful, and raises the question as to whether there is a better alternative. Such an alternative is provided by the concept of a *regenerating code* introduced in the pioneering paper by Dimakis et al. [4].

## A. Regenerating Codes

In the regeneration framework introduced in [4], codes whose symbol alphabet is a vector over $\mathbb{F}_q$, i.e., an element of $\mathbb{F}_q^\alpha$ for some parameter $\alpha > 1$ are employed. As with RS codes, each node still continues to store a single code symbol. However, given the vector nature of the code-symbol alphabet, we may equivalently, regard each node as storing a collection of $\alpha$ symbols, each symbol drawn from $\mathbb{F}_q$. Under this setup, it is clear that while maintaining linearity over $\mathbb{F}_q$, it is possible for an individual node to transfer a fraction of the data stored within the node.

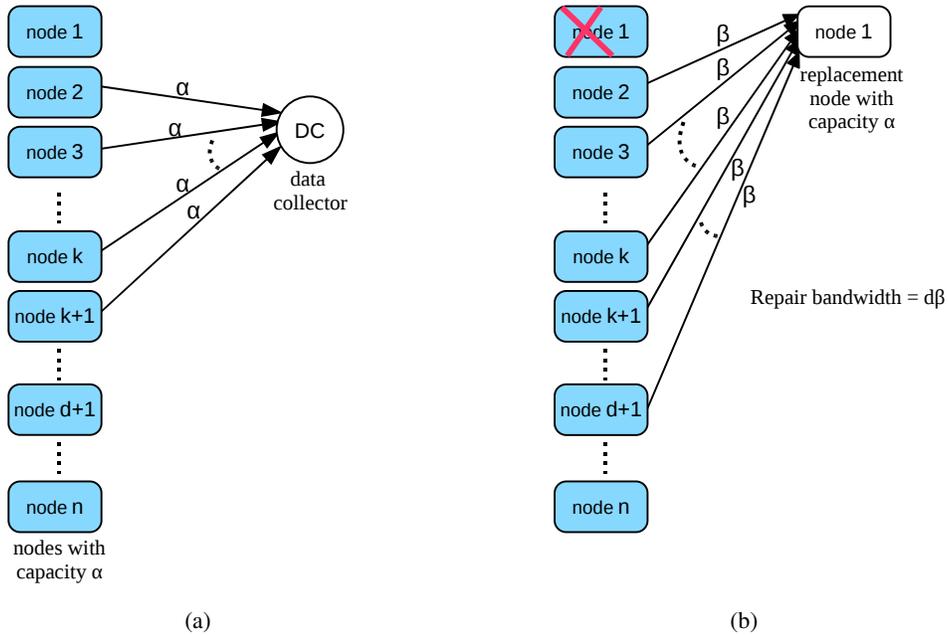

Fig. 1: The regenerating codes setup: (a) data-reconstruction, and (b) repair of a failed node.

Apart from this new parameter $\alpha$, two other parameters $(d, \beta)$ are associated with the regenerating framework introduced in [4]. A replacement node is permitted to connect to an arbitrary subset of $d$ nodes out of the remaining $(n-1)$ nodes while downloading $\beta \leq \alpha$ symbols from each node. The $d$ nodes helping in the repair of a failed node are termed as *helper nodes*. The total amount $d\beta$ of data downloaded for repair purposes is termed the *repair bandwidth*. Thus under this framework, apart from the field size $q$, we have

$$\{ [n, \ k, \ d], \ (\alpha, \beta, \ B) \}$$

as the parameter set. The corresponding codes are called *regenerating codes*. Typically, with a regenerating code, the average repair bandwidth $d\beta$ is small compared to the size of the file $B$. Fig. 1a and Fig. 1b illustrate reconstruction and node repair respectively, while also depicting the relevant parameters.

Note that the parameters $k$ and $d$ are the minimum values under which reconstruction and repair can always be guaranteed. This restricts the range of $d$ to

$$k \leq d \leq n-1, \tag{1}$$

for, if the repair parameter $d$ were less than the reconstruction parameter $k$, this would imply that one could in fact reconstruct the data by connecting to $d$ nodes, thereby contradicting the minimality of $k$.

## B. Exact versus Functional Repair

Under the notion of *functional repair* introduced in [4], a failed node is replaced by a node that is functionally equivalent, i.e., following replacement, the resulting network of $n$ nodes must continue to possess the reconstruction and repair properties. In contrast, under *exact-repair*, introduced subsequently in [7], [8], a replacement node is required to store exactly the same data as was stored in the failed node. Hence, there is no change in the coefficients of a replaced node under exact-repair. This obviates additional communication overheads during the repair operation, and also avoids re-tuning of the reconstruction and repair algorithms. Thus exact-repair greatly simplifies system implementation and is of considerable practical interest.

We use the term *exact-repair code* to denote a regenerating code that is the capable of performing exact-repair of any failed node.

## C. The Storage-Repair Bandwidth Tradeoff

A major result in the field of regenerating codes is the proof in [5] that uses the cut-set bound of network coding to establish that the parameters of a regenerating code must necessarily satisfy [1]:

$$B \leq \sum_{i=0}^{k-1} \min\{\alpha, (d-i)\beta\}. \tag{2}$$

It is desirable to minimize both $\alpha$ as well as $\beta$ since, minimizing $\alpha$ results in a minimum storage solution, while minimizing $\beta$ results in a storage solution that minimizes repair bandwidth. As can be deduced from (2), it is not possible to minimize both $\alpha$ and $\beta$ simultaneously and thus there exists a tradeoff between the choices of the parameters $\alpha$ and $\beta$ and this tradeoff is termed as the Storage-Repair Bandwidth Tradeoff, or more simply as the storage-bandwidth tradeoff. In Fig. 2, the tradeoff is plotted for $B = 27000$ symbols for a system with $k = 10$, $d = 18$ and some $n > 18$.

For fixed values of $B$ and $[n, \ k, \ d]$, a regenerating code is said to be *optimal*, if the parameters $(\alpha, \ \beta)$ are such that:

1) equality holds in equation (2), i.e.,

$$B = \sum_{i=0}^{k-1} \min\{\alpha, (d-i)\beta\}, \tag{3}$$

---

[1]This bound on the message size $B$ is originally derived in [5] using the principles of network coding. An information-theoretic derivation is presented in Section IV-B in the present paper.

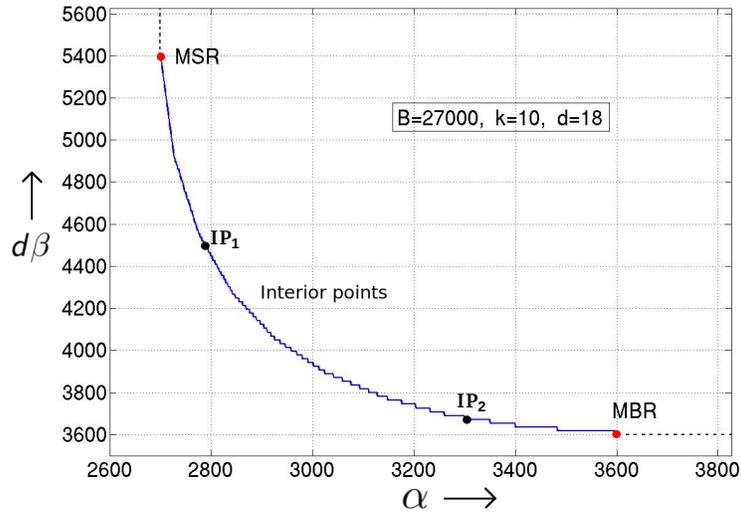

Fig. 2: The storage-bandwidth tradeoff curve: storage space $\alpha$ vs repair bandwidth $d\beta$. Also depicted are the two end points MBR and MSR, and two interior points IP$_1$ and IP$_2$.

and

2) if either $\alpha$ or $\beta$ is decreased, equation (3) fails to hold.

The parameters $(\alpha, \beta)$ of any optimal regenerating code are said to *lie on the storage-bandwidth tradeoff*.

Observe that when $\alpha < (d - (k-1))\beta$, the parameter $\beta$ can be decreased without violating (3). Hence the parameters of an optimal regenerating code must necessarily satisfy

$$\alpha \geq (d - k + 1)\beta. \tag{4}$$

Thus, the case of

$$\alpha = (d - k + 1)\beta \tag{5}$$

is one of the extreme points of the tradeoff called the Minimum Storage Regenerating (MSR) point. From equation (3), we see that the parameters at the MSR point satisfy

$$B = k\alpha. \tag{6}$$

On the other hand, if $\beta < \frac{\alpha}{d}$ (i.e., if $\alpha > d\beta$), the parameter $\alpha$ can be decreased without violating (3). Hence the parameters of an optimal regenerating code must necessarily satisfy

$$\beta \geq \frac{\alpha}{d}. \tag{7}$$

The case when $\beta = \frac{\alpha}{d}$ i.e.,

$$\alpha = d\beta \tag{8}$$

is the other extreme point of the tradeoff called the Minimum Bandwidth Regenerating (MBR) point. From equation (3) we see that the parameters at the MBR point satisfy

$$B = \left( kd - \frac{k(k-1)}{2} \right) \beta. \tag{9}$$

It can be inferred from equations (4) and (7) that an optimal regenerating code must have the parameter $\alpha$ lying in the range

$$(d - k + 1)\beta \leq \alpha \leq d\beta. \tag{10}$$

Points on the tradeoff other than the two extreme (MSR and MBR) points, have the parameter $\alpha$ lying strictly within this range: $(d - k + 1)\beta < \alpha < d\beta$. These points are hence referred to as the *interior points* on the tradeoff curve.

Note that the situation when $k = 1$ results in $B = \alpha = d\beta$ which can be satisfied trivially by a repetition code. Thus, we assume $k > 1$ through the rest of the paper.

The storage-bandwidth tradeoff was derived in [5] for the case of functional repair and was shown to be tight in [5], [6]. Clearly, the bound continues to hold even under the exact-repair setting, since exact-repair is an instance of a functional-repair. However, the achievability of this bound under an exact-repair requirement has remained an open problem, and will be addressed in the present paper.

### D. Summary of the Results in This Paper

The first main result of the present paper is an explicit construction of exact-MBR codes for the case $d = n - 1$, that has a simple graphical description. An interesting, and potentially useful, aspect of this construction is its *repair-by-transfer* property where a failed node is repaired by simple transfer of data without need for any computation at either the helper nodes or the replacement node. Furthermore, when specialized to the parameter set $[n, \ k = n - 2, \ d = n - 1]$, encoding of the code as well as data-reconstruction can be accomplished using XOR operations alone. These properties give the code practical appeal [2].

The second main result of the present paper answers an open problem regarding the achievability of the storage-bandwidth tradeoff under exact-repair, at the interior points. First, a set of properties required to be satisfied by an exact-repair code are derived; in particular, a set of constraining scenarios, which we call 'helper node pooling' scenarios, are identified. Subsequently, the non-achievability of the interior points on the storage-bandwidth tradeoff under exact-repair is established, with the possible exception of points within the immediate vicinity of the MSR point.

---

[2] An animated video of an example of this code is available in [24].

*E. Organization*

The paper is organized as follows. A brief overview of related literature is provided in Section II. Section III contains the exact-MBR code construction. A set of properties that any exact-repair code must necessarily satisfy are provided in Section IV, which are then used to establish the non-achievability of the storage-bandwidth tradeoff for exact-repair at nearly all interior points. Section V presents conclusions.

## II. Related Work

The concept of regenerating codes, introduced in [4], [5], permit storage nodes to store more than the minimal $B/k$ units of data in order to reduce the repair bandwidth. Several distributed systems are analyzed, and estimates of the mean node availability in such systems are obtained. Using these values, the substantial performance gains offered by regenerating codes in terms of bandwidth savings are demonstrated. The problem of minimizing repair bandwidth for *functional* repair of nodes is formulated as a multicast network-coding problem in a network having an infinite number of nodes. A cut-set lower bound on the repair bandwidth is derived. Coding schemes achieving this bound are presented in [5], [6] which however are non-explicit. These schemes require large field size and the repair and reconstruction algorithms are also of high complexity.

The notion of exact-repair is introduced independently in [7] and [8]. In [7], the MSR point is shown to be achievable under exact-repair for the parameters $[n, \ k = 2, \ d = n - 1]$. The proposed coding scheme uses the concept of interference alignment. Even here, the constructions are not explicit, and have large complexity and field-size requirement.

The first explicit construction of regenerating codes appears in [8]. An explicit MSR code is constructed here, for $[n, \ k, \ d = k + 1]$ (see also [11]). A computer search for exact-repair MSR codes for the parameter set $[n = 5, k = 3, \ d = 4]$ is carried out in [9], and for this set of parameters, codes for several values of field size are obtained.

A slightly different setting, from the exact-repair situation is considered in [13], where optimal MDS codes are given for the parameters $[n > 2k, \ k, \ d = k + 1]$. Again, the schemes given here are non-explicit, and have high complexity and large field-size requirement.

An explicit code structure at the MSR point, that guarantees reconstruction and exact-repair of the systematic nodes is provided in [10], for parameters $[n \geq 2k, \ k, \ d = n - 1]$. This code makes use of interference alignment, and is termed as the 'MISER' code in the journal-submission version [11] of [10]. Following the initial submission of [10], it is shown in [12] that for this set of parameters, the code introduced in [10] can perform exact-repair of parity nodes as well, and such repair schemes are indeed designed and presented. The impossibility of constructing linear, scalar (i.e., $\beta = 1$) exact-MSR codes when $d < 2k - 3$ is shown in [10], [11]. On the other hand, in the limiting case of $B$ (and hence $\alpha$ and $\beta$) approaching infinity, the MSR point is shown to be achievable under exact-repair for all $[n,\ k,\ d]$ in [14], [15].

A general framework, termed the *Product-Matrix* framework, that enables construction for a wide range of parameters is introduced in [16]. Explicit codes at the MBR point for all values of the parameters $[n,\ k,\ d]$ and at the MSR point for the parameter set $[n,\ k,\ d \geq 2k-2]$ are constructed in this framework. Also contained in the paper is a simpler description of the MISER code in the Product-Matrix framework.

There has also been some work in the literature that consider slightly different models for distributed storage systems. Codes for more relaxed settings with respect to the data recovery/repair requirements are presented in [17], [18]. The papers [19]–[22] provide alternative frameworks for regenerating codes that introduce additional parameters for the system; tradeoffs between storage and repair are derived in each of the papers for the functional repair scenario.

Following the initial presentation of the exact-MBR codes performing repair-by-transfer in [8] (described in Section III in the present paper), El Rouayheb et al. [23] use the graphical approach to construct codes for a larger set of parameters, in a somewhat relaxed setting where a replacement node can connect to only certain fixed subsets of $d$ nodes for repair [3]. They also provide upper bounds on the storage capacity of such systems.

## III. Explicit Exact-MBR Code for $d = n-1$ with Repair-By-Transfer

In this section, we provide an explicit construction of exact-MBR codes wherein the parameter $d$ takes the largest permissible value of $n-1$ [4]. These codes are capable of performing exact-repair of any failed node in such a way that the repair process is accomplished with mere transfer of data and without need for any arithmetic operations either at the helper nodes or at the replacement node. This property makes the code practically appealing.

First we present a brief overview of the Minimum Bandwidth Regenerating point on the storage-bandwidth tradeoff and the concept of *striping* of data which will be used in the code construction.

### A. *MBR Point Parameters*

The MBR point is as extreme point on the storage-bandwidth tradeoff that corresponds to the least possible repair bandwidth. As previously discussed in Section I-C, the parameters $\alpha$ and $\beta$ for the MBR

---

[3]In this work, the authors refer to the repair-by-transfer property of codes in [8] as *uncoded-repair*.

[4]It can be inferred from the storage-bandwidth tradeoff (3) that, for fixed values of the parameters $n,\ k,\ \alpha$ and $B$, the repair bandwidth $d\beta$ decreases with increase in $d$.

point satisfy

$$\begin{aligned}
\alpha &= d\beta \\
B &= \left(kd - \frac{k(k-1)}{2}\right)\beta.
\end{aligned} \tag{11}$$

Thus at the MBR point, a replacement node downloads no more than the number of symbols it eventually stores.

At this point, we briefly digress to analyze a particular relation between the parameters $(\alpha, \beta, B)$ of a regenerating code, that will aid in simplifying the system implementation of the code.

## B. The Striping of Data

Given a set of parameters $(B, \alpha, \beta)$ satisfying the tradeoff (2) with equality, the parameters $(\alpha' = \delta\alpha, \beta' = \delta\beta, B' = \delta B)$ for any positive integer $\delta$ also satisfy the tradeoff (2) with equality. Thus, for some $[n, k, d]$, an optimal regenerating code for $(\alpha', \beta', B')$ can be obtained easily by dividing the $B' = \delta B$ message symbols into $\delta$ groups of $B$ symbols each, and applying the optimal $(B, \alpha, \beta)$ code to each group independently. In particular, if one can construct an $[n, k, d]$ MBR code with $\beta = 1$, then one can construct an $[n, k, d]$ MBR code for any larger integer value of $\beta$ as well.

From a practical standpoint, a code with smaller $\beta$ will involve manipulating a smaller number of message symbols and hence will in general, be of lesser complexity. For these reasons, in the present paper we design codes for the simplest case of $\beta = 1$. In this situation, the values of $\alpha$ and $B$ at the MBR point are given by

$$\begin{aligned}
\alpha &= d \\
B &= kd - \binom{k}{2}.
\end{aligned} \tag{12}$$

## C. An Example Code

The example deals with the parameter set $[n = 5, k = 3, d = 4]$, $\beta = 1$ which from equation (12) gives $\alpha = 4$ and $B = 9$. Let the 9 message symbols be denoted by $\{m_i | 1 \le i \le 9\}$.

*Encoding:* Let $\mathcal{C}$ be a $[10, 9]$-MDS code, for example, $\mathcal{C}$ could be a single parity check code of length 10. Encode the 9 message symbols using the code $\mathcal{C}$, and let the 10 coded symbols be denoted by $\{c_i | 1 \le i \le 10\}$. Our code construction can be visualized via a fully connected graph (see Fig. 3) on 5 nodes, each representing a distinct node in the network. To each of the 10 edges in the graph, we assign a distinct symbol from the set of code symbols $\{c_i | 1 \le i \le 10\}$ (in any particular order).

In our construction, each storage node stores the 4 symbols assigned to the four edges incident on the node, as shown in Fig. 3.

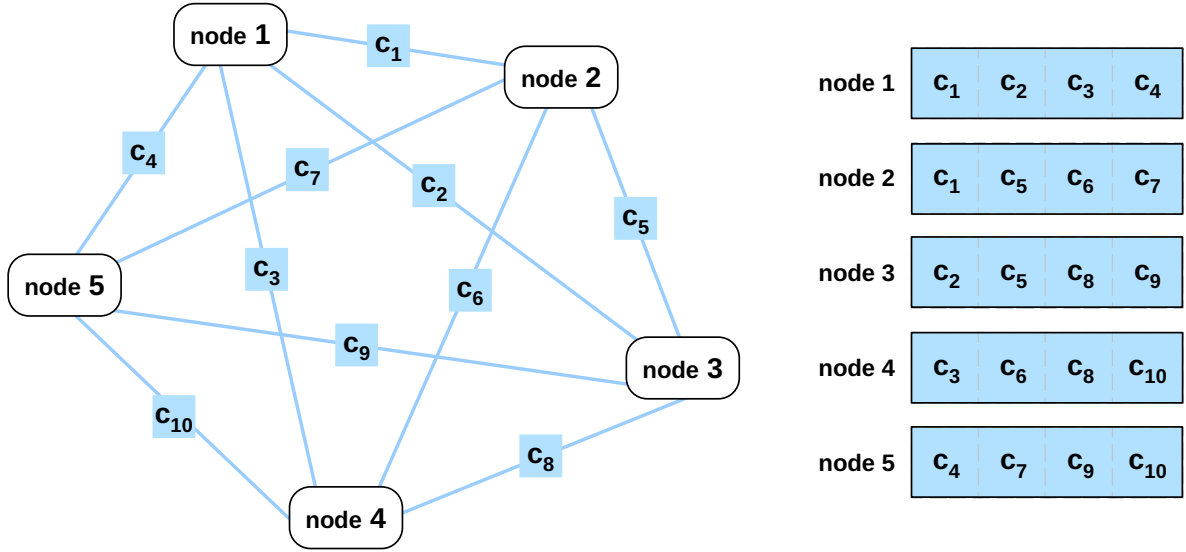

Fig. 3: Graphical representation of an exact-repair code for the MBR point with $[n = 5, \ d = 4, \ k = 3]$.

*Data-Reconstruction:* Supposing a data-collector connects to nodes 2, 3 and 4. The data-collector then recovers $3 \times 4 = 12$ coded symbols of which $12 - \binom{3}{2} = 9$ are distinct. Since the code $\mathcal{C}$ is an $[10, \ 9]$-MDS code by construction, the data-collector can recover the 9 message symbols from these 9 coded symbols.

*Repair-by-transfer:* Suppose node 3 fails, and is replaced by an empty replacement node. Under our construction, each of the four remaining nodes pass the symbol assigned to the edge that it has in common with node 3, i.e., nodes 1, 2, 4 and 5 pass on the symbols $c_2$, $c_5$, $c_8$ and $c_9$ respectively to the replacement node (from Fig. 3). But these are precisely the four symbols that were stored in node 3 prior to failure, and hence node 3 is exactly repaired. Note that the repair is accomplished by mere transfer of data and no arithmetic operations are required either at the helper nodes or at the replacement node. Hence we term this as *repair-by-transfer*, which is potentially useful in practical systems.

## D. Code Construction for the General Set of Parameters $[n, \ k, \ d = n - 1]$

As discussed previously, the code is constructed for the case of $\beta = 1$, and codes for any higher value of $\beta$ can be obtained via concatenation. The construction follows along the lines similar to the example provided in the previous section.

Let the $B$ message symbols be denoted by $\{m_i | 1 \leq i \leq B\}$.

*Encoding:* Let $\mathcal{C}$ be an $[\binom{n}{2}, \ B]$-MDS code. Encode the $B$ message symbols using the code $\mathcal{C}$, and let the $\binom{n}{2}$ coded symbols be denoted by $\{c_i | 1 \leq i \leq \binom{n}{2}\}$. As in the example, we can visualize the code construction via a fully connected graph on $n$ nodes, each representing a distinct node in the network. To each of the $\binom{n}{2}$ edges in the graph, we assign a distinct code symbol from the set $\{c_i | 1 \leq i \leq \binom{n}{2}\}$ (in any particular order).

In our construction, each storage node stores the $\alpha$ $(= n - 1)$ symbols assigned to the $(n-1)$ edges incident on the node. Thus, each symbol in $\{c_i | 1 \leq i \leq \binom{n}{2}\}$ is stored in precisely two nodes.

The following theorems establish the properties of data-reconstruction and repair-by-transfer.

*Theorem 1 (Data-Reconstruction):* A data-collector can recover all the $B$ message symbols by connecting to any subset of $k$ nodes.

*Proof:* The data-collector connects to a subset of $k$ nodes in the network, and recovers the $k\alpha$ $(= kd)$ code symbols stored in these nodes. Since every pair of nodes has exactly one symbol in common, there are $\binom{k}{2}$ redundant symbols among these $kd$ code symbols. Thus, the data-collector has access to $kd - \binom{k}{2} = B$ distinct code symbols from the set $\{c_i | 1 \leq i \leq \binom{n}{2}\}$. Since the code $\mathcal{C}$ is an $[\binom{n}{2}, B]$-MDS code by construction, the data-collector can recover the $B$ message symbols $\{m_i | 1 \leq i \leq B\}$ from these $B$ code symbols. ∎

Note that in our construction, the data-reconstruction (decoding) procedure is identical to that of decoding an MDS code over an erasure channel.

*Theorem 2 (Repair-By-Transfer):* Exact-repair of any failed node can be achieved by connecting to the remaining $(n-1)$ nodes and downloading the minimum possible data. Furthermore, the process involves mere transfer of data and does not require any arithmetic operations.

*Proof:* On failure of a storage node, the replacement node connects to the $(n-1)$ remaining nodes. Each of the remaining nodes passes to the replacement node, the symbol assigned to the edge that it has in common with the failed node. By construction, these symbols are precisely the $(n-1) = \alpha$ symbols that were stored in the node prior to failure. Thus, the replacement node simply stores these symbols, completing the process of exact-repair. Clearly, the repair process does not require any arithmetic operation either at the helper nodes or at the replacement node. ∎

### E. Size of the Finite Field

The sole constraint on the field size required in the code construction arises from the need for the existence of an $[\binom{n}{2}, B]$ MDS code. The existence of doubly extended Reed-Solomon codes tells us that a field size of $\binom{n}{2} - 1$ will suffice.

*Remark 1:* The example employed a single parity check code as its MDS code. This has practical appeal since all operations can be carried out in binary field using only XORs. We note from the above that a single parity check code will suffice whenever $k = n - 2$, which from equation (12) gives $\binom{n}{2} = B + 1$.

## IV. Non-existence of Exact-Repair codes achieving the Interior Points on the Storage-Bandwidth Tradeoff

We now move on to the second main result of the paper which proves the non-achievability of the interior points on the storage-bandwidth tradeoff under exact-repair. Originally, the storage-bandwidth tradeoff was derived for the case of functional-repair. However, given the clear advantages of exact-repair, much of the work in the field of regenerating codes has been dedicated to exact-repair.

In particular, the two extreme points of the tradeoff (MBR and MSR) have been much studied, and the achievability of the tradeoff for exact-repair at the extreme points has been characterized to a large extent. Interference Alignment based explicit constructions in [10]–[12] achieve the cut-set bound at the MSR point for $[n = d + 1, \ k, \ d \geq 2k - 1]$. Product-Matrix framework based exact-repair code constructions in [16] cover all possible values of the parameters $[n, \ k, \ d]$ at the MBR point, and all $[n, \ k, \ d \geq 2k - 2]$ at the MSR point, thus proving the tightness of the storage-bandwidth tradeoff for these regimes. Further, in [10], [11], the authors show that there exist no linear, scalar (i.e., with $\beta = 1$) MSR codes performing exact-repair when $d < 2k - 3$. In [14], [15], the authors show that in the asymptotic case of $\alpha$, $\beta$ and $B$ tending to infinity, the MSR point can be achieved under exact-repair for all possible $[n, \ k, \ d]$.

However, the tightness of the storage-bandwidth tradeoff under exact-repair has remained open for the interior points. In the present section, we address this issue by proving the impossibility of constructing codes performing exact-repair at almost all interior points on the storage-bandwidth tradeoff. Furthermore, we identify scenarios, which we term as helper node pooling scenarios, that overconstrain the system.

This section is organized as follows. First, an information-theoretic perspective for regenerating codes is presented in Section IV-A, which is used throughout the present section. Next, the storage-bandwidth tradeoff (2) is re-derived information-theoretically in Section IV-B. Section IV-C presents a convenient means of representing every point on the storage-bandwidth tradeoff as a function of parameters $\alpha$ and $\beta$. Following this, a set of conditions imposed by the cut-set bound on the amount of information stored and passed by the nodes are derived in Section IV-D. In addition, 'helper node pooling' scenarios are defined in this section. These properties are then utilized in Section IV-E to establish non-achievability of the interior points on the storage-bandwidth tradeoff for exact-repair. Finally, in Section IV-F, an achievable curve is plotted via a simple storage-space sharing scheme between existing codes for the two extreme points.

### A. Notation

While in earlier sections we worked with individual symbols from a certain alphabet, in keeping with the information-theoretic approach of the current section, we treat the message symbols as well as the data stored and passed by the nodes as random variables.

Under this information-theoretic perspective, the nodes in the network store data pertaining to a source (message) $M$, whose entropy is $B$, i.e.,

$$H(M) = B. \tag{13}$$

Next, we introduce the random variables, pertaining to the data stored in the nodes and the data passed by nodes for data-recovery and repair purposes.

Let $W_\ell$ denote the random variable corresponding to the data stored in node $\ell$ $(1 \leq \ell \leq n)$. We will assume that each storage node has a storage capacity of $\alpha$ and is hence incapable of storing variables whose entropy are greater than $\alpha$, thus

$$H(W_\ell) \leq \alpha. \tag{14}$$

Consider exact-repair of node $\ell$ using a set $\mathcal{D}$ of $d$ helper nodes, and let node $m \in \mathcal{D}$. In this situation, we denote the random variable corresponding to the data passed by the helper node $m$ to aid in the repair of node $\ell$ by $_\mathcal{D}S_m^\ell$. We also assume that the data pipes used for repair have capacity $\beta$ and hence are incapable of carrying variables whose entropy are greater than $\beta$, i.e.,

$$H(_\mathcal{D}S_m^\ell) \leq \beta. \tag{15}$$

Both equation (14) and equation (15) are in keeping with the original setting where each node had the capacity to store $\alpha$ symbols and each data pipe used in repair had the capacity to carry $\beta$ symbols.

Note that since repair is exact, the random variables $W_\ell$ and $_\mathcal{D}S_m^\ell$ are invariant with time, i.e., they remain constant irrespective of the sequence of failures and repairs that occur in the system [5].

A little thought shows that the reconstruction and exact-repair requirements can be stated information-theoretically as follows.

(i) From the reconstruction property required of a regenerating code, it must be that, for every subset of $k$ storage nodes: $\{\ell_i \mid 1 \leq i \leq k\}$, we need

$$H\left(M \,\middle|\, \{W_{\ell_i}\}_{i=1}^k\right) = 0. \tag{16}$$

(ii) Similarly, the exact-repair property requirement leads to the condition that for every node $\ell$ $(1 \leq \ell \leq n)$

$$H\left(W_\ell \,\middle|\, \{_\mathcal{D}S_m^\ell\}_{m \in \mathcal{D}}\right) = 0. \tag{17}$$

where $\mathcal{D}$ represents the set of $d$ helper nodes participating in the exact-repair of node $\ell$.

In the sequel, for simplicity, we will drop the left subscript $\mathcal{D}$ and the set of $d$ nodes participating in the repair process will be clear from the context. Furthermore, since a node can pass only a function

---

of what is stored in the node, it follows that, for any node $m$,

$$H\left(S_m^\ell \,\middle|\, W_m\right) = 0. \tag{18}$$

*Remark 2:* Throughout this section, we will assume that all the random variables are functions of the message $M$. This is without loss of generality since one can always assume a genie that reveals all the extraneous sources of randomness to every entity in the system, and this would still retain the necessity of the properties proved here. However, for convenience, we do not indicate the dependence in the notation. Thus, we have

$$H(W_\ell|M) = 0 \quad \text{and} \quad H(S_m^\ell|M) = 0. \tag{19}$$

Now, from equations (13) and (19), one can rewrite the reconstruction property in (16) as

$$H\left(\{W_{\ell_i}\}_{i=1}^k\right) = B. \tag{20}$$

Next, we set up notation to denote certain sets of random variables which will be used frequently. Let $\mathcal{A}$ denote a collection of storage nodes. Then the set of random variables corresponding to the data stored in the nodes in $\mathcal{A}$ is denoted by

$$W_{\mathcal{A}} \triangleq \{W_i\}_{i \in \mathcal{A}}. \tag{21}$$

Further, define $[m]$ as the set of numbers $\{1, \ldots, m\}$ for some positive integer $m$, and denote

$$W_{[m]} \triangleq \{W_i\}_{i=1}^m. \tag{22}$$

Note that the notation $[0]$ will correspond to an empty set.

The random variables corresponding to the data passed for repair may be grouped in two ways. Denote the collection of random variables passed by nodes in set $\mathcal{A}$ to assist in the repair of a particular failed node $\ell$ by

$$S_{\mathcal{A}}^\ell \triangleq \left\{S_m^\ell\right\}_{m \in \mathcal{A}}. \tag{23}$$

On the other hand, across the instances of failure of every node in set $\mathcal{A}$, the collection of variables passed by a particular helper node $m$ to each of these nodes respectively is denoted by

$$S_m^{\mathcal{A}} \triangleq \left\{S_m^\ell\right\}_{\ell \in \mathcal{A}}. \tag{24}$$

Note that in both the cases above, the other helper nodes participating in the repair process will be clear from the context.

*B. Information-Theoretic Derivation of the Tradeoff*

We now present an information-theoretic derivation of the storage-bandwidth tradeoff (2), since it is convenient to remain in the information-theoretic domain throughout this section. The results established in this subsection are derived only for the case of exact-repair. The extensions to the case of functional-repair are straightforward and are explained subsequently.

The following lemma establishes a relation between the information stored in various nodes.

*Lemma 3:* For an arbitrary node $\ell$, and an arbitrary subset $\mathcal{A}$ consisting of $a$ $(0 \leq a \leq d)$ nodes such that $\ell \notin \mathcal{A}$,

$$H\left(W_\ell \left| S_\mathcal{A}^\ell \right.\right) \leq \min(\alpha, \ (d-a)\beta) \tag{25}$$

and hence

$$H\left(W_\ell \left| W_\mathcal{A} \right.\right) \leq \min(\alpha, \ (d-a)\beta). \tag{26}$$

*Proof:* Consider exact-repair of node $\ell$ by connecting to the $a$ nodes in set $\mathcal{A}$ and $(d-a)$ other arbitrary nodes. Denote the set of these $(d-a)$ helper nodes by $\mathcal{B}$. Then, the exact-repair of node $\ell$ requires

$$0 \ = \ H\left(W_\ell \left| S_\mathcal{A}^\ell, \ S_\mathcal{B}^\ell \right.\right) \tag{27}$$

$$= \ H\left(W_\ell \left| S_\mathcal{A}^\ell \right.\right) - I\left(W_\ell; S_\mathcal{B}^\ell \left| S_\mathcal{A}^\ell \right.\right) \tag{28}$$

$$\geq \ H\left(W_\ell \left| S_\mathcal{A}^\ell \right.\right) - H\left(S_\mathcal{B}^\ell\right) \tag{29}$$

$$\geq \ H\left(W_\ell \left| S_\mathcal{A}^\ell \right.\right) - (d-a)\beta \tag{30}$$

$$\geq \ H\left(W_\ell \left| W_\mathcal{A} \right.\right) - (d-a)\beta \ , \tag{31}$$

where equation (30) follows since each of the $(d-a)$ helper nodes in $\mathcal{B}$ can pass at most $\beta$ units of information to the replacement node, and equation (31) is a result of the fact that a node can only pass a function of what it stores, i.e., $H(S_m^\ell | W_m) = 0$ for all nodes $m \in \mathcal{A}$. Equation (31), coupled with the constraint on the storage capacity of the nodes (i.e., $H(W_\ell) \leq \alpha$) leads to the desired result. ∎

*Remark 3 (The case of functional-repair):* In the case of functional repair, the data stored in a node after repair need not be identical to that stored in it prior to failure. This makes the corresponding random variable a function of time. In this scenario, the lemma above applies when $W_\ell$ is the random variable corresponding to the data stored in node $\ell$ after being repaired with the help of $a$ nodes in $\mathcal{A}$ and $(d-a)$ other arbitrary nodes.

The next theorem gives a simple derivation of the storage-bandwidth tradeoff (2) for exact-repair from an information-theoretic perspective.

*Theorem 4:* Any $[n, k, d]$, $(B, \alpha, \beta)$ regenerating code must necessarily satisfy

$$B \leq \sum_{\ell=0}^{k-1} \min\{\alpha, (d-\ell)\beta\}. \tag{32}$$

*Proof:* The reconstruction property (recall equation (20)) requires

$$B = H\left(W_{[k]}\right) \tag{33}$$

$$= \sum_{\ell=0}^{k-1} H\left(W_{\ell+1} \,\big|\, W_{[\ell]}\right) \tag{34}$$

$$\leq \sum_{\ell=0}^{k-1} \min\{\alpha, (d-\ell)\beta\}, \tag{35}$$

where equation (35) follows from Lemma 3. ∎

*Remark 4 (The case of functional-repair):* The above theorem holds for the case of functional-repair as well. Here, a sequence of failures and repairs is considered, starting from node 2 through node $k$ (in this order). The $d$ nodes assisting node $\ell + 1$ ($1 \leq \ell \leq k-1$) in its repair include the $\ell$ nodes in set $[\ell]$, and $W_\ell$ is the random variable corresponding to the data stored in node $\ell$ after its repair.

Next, we present a convenient way to represent all points on the storage-bandwidth tradeoff in terms of $\alpha$ and $\beta$.

## C. Representation for the Points on the Tradeoff

As previously discussed in Section I-C, for any $[n, k, d]$ optimal regenerating code, the parameter $\alpha$ lies in the range

$$(d - (k-1))\beta \ \leq \ \alpha \ \leq \ d\beta. \tag{36}$$

Based on this, for a given value of $(\alpha, \beta)$, we depict the relation between the two parameters as

$$\alpha = (d-p)\beta - \theta, \tag{37}$$

for some $p$ and $\theta$ with $p \in \{0, \ldots, k-1\}$ and $\theta \in [0, \beta)$. Note that the range of $\alpha$ in (36) implies that for $p = k - 1$, it must be that $\theta = 0$.

The storage-bandwidth tradeoff can thus be partitioned into the two end-points and a middle region:

1) The MSR point: $p = k - 1$ (which implies $\theta = 0$)
2) The MBR point: $p = 0$, $\theta = 0$
3) The Interior points: $p \in \{0, \ldots, k-2\}, \theta \in [0, \beta)$ except $\{p = 0, \theta = 0\}$.

For instance, the values of $(\alpha, \beta, p, \theta)$ at the four points depicted on the storage-bandwidth tradeoff

in Fig. 2 are

$$
\begin{aligned}
\text{MSR} &: \ (2700,\ 300,\ 9,\ 0), \\
\text{IP}_1 &: \ (2786,\ 250,\ 6,\ 214), \\
\text{IP}_2 &: \ (3300,\ 204,\ 1,\ 168), \\
\text{MBR} &: \ (3600,\ 200,\ 0,\ 0).
\end{aligned}
$$

### D. Properties of Exact-Repair Codes

We now present a set of properties that any exact-repair code with parameters satisfying the storage-bandwidth tradeoff with equality (3), must necessarily possess. These properties pertain to the random variables stored by the nodes and those passed for exact-repair. The proofs of these properties are relegated to Appendix A.

The first two properties provide insights pertaining to data stored in the nodes, and the subsequent properties provide insights about the data passed for repair.

*Property 1 (Entropy of Data Stored):* For an arbitrary storage node $\ell$,

$$H\left(W_\ell\right) = \alpha. \tag{38}$$

*Property 2 (Mutual information among the nodes):* For a set $\mathcal{A}$ comprising of an arbitrary collection of $a$ nodes, and an arbitrary node $\ell \notin \mathcal{A}$,

$$
I\left(W_\ell; W_\mathcal{A}\right) = \left\{
\begin{array}{ll}
0 & a \leq p \\
(a-p)\beta - \theta & p < a < k \\
\alpha & a \geq k \ .
\end{array}
\right. \tag{39}
$$

Note that in Property 2 above, a threshold effect manifests itself twice in the mutual information, the first threshold occurring at $a = p + 1$ and the second at $a = k$. This is illustrated in Fig. 4a. This is a phenomenon similar (albeit more complex) to the single threshold effect in MDS codes where a code symbol has zero mutual information with upto $(k-1)$ other code symbols and has mutual information equal to its entropy with any $k$ or more other code symbols.

Further, note that with respect to the amount of mutual information between a single node and a bunch of nodes, the two extreme points behave differently as compared to the interior points. This is depicted in Fig. 4b and Fig. 4c. As it turns out, it is this difference in behaviour that leads to the non-existence for the interior points.

*Corollary 5:* Consider a set $\mathcal{A}$ comprising of an arbitrary collection of $a < k$ nodes. In the situation where the set of $d$ helper nodes assisting in the repair of an arbitrary node $\ell \notin \mathcal{A}$ include the $a$ nodes in $\mathcal{A}$, it must be that

$$H(W_\ell | S_\mathcal{A}^\ell) = \min\{\alpha, (d-a)\beta\}. \tag{40}$$

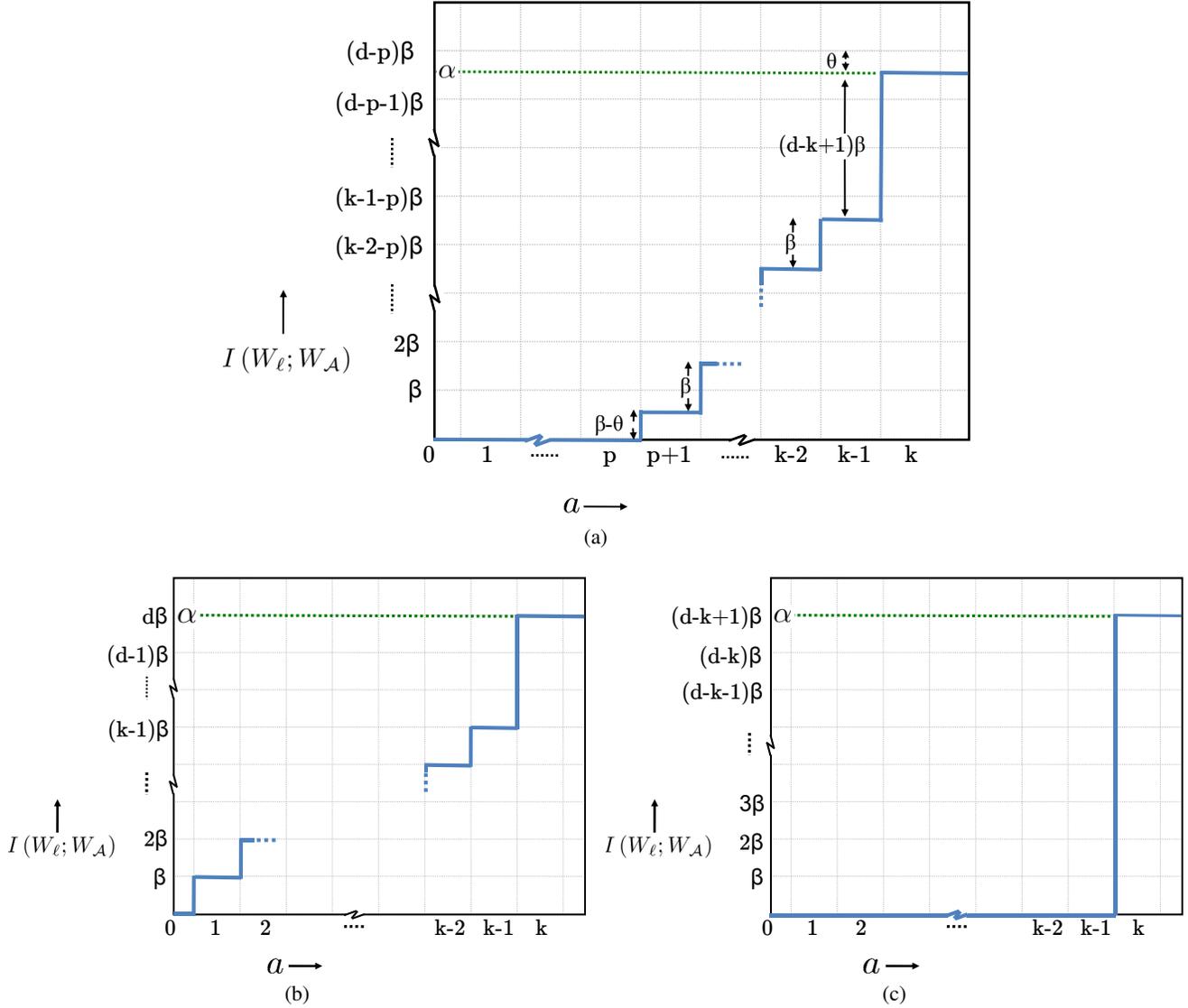

Fig. 4: The amount of information that a node $\ell$ has in common with a set $A$ of $a$ other arbitrary nodes (Property 2): for (a) a general point on the tradeoff, (b) the MBR point, and (c) the MSR point.

*Property 3 (Entropy of Data Passed):* In the situation where node $m$ is an arbitrary helper node assisting in the repair of a second arbitrary node $\ell$, it must be that

$$H\left(S_m^\ell\right) = \beta, \tag{41}$$

irrespective of the identity of the $(d-1)$ other helper nodes.

*Helper Node Pooling:* Regenerating codes permit a failed node to choose an arbitrary set of $d$ remaining nodes to aid in its repair. In particular, this includes situations where nodes form a pool and help each other in the repair process. More formally, consider a set $\mathcal{F}$ consisting of a collection of $f \leq (d+1)$ nodes, and a subset $\mathcal{R}$ of the set $\mathcal{F}$ consisting of $r$ nodes. We refer to 'helper node pooling' as a scenario where on failure of any node $\ell \in \mathcal{R}$, the $d$ helper nodes assisting in its repair include the $(f-1)$ remaining nodes in $\mathcal{F}$. We denote the remaining $(d-(f-1))$ arbitrary helper nodes assisting in the repair of node $\ell$ by $\mathcal{V}(\ell)$ [6]. The helper node pooling scenario is illustrated in Fig. 5.

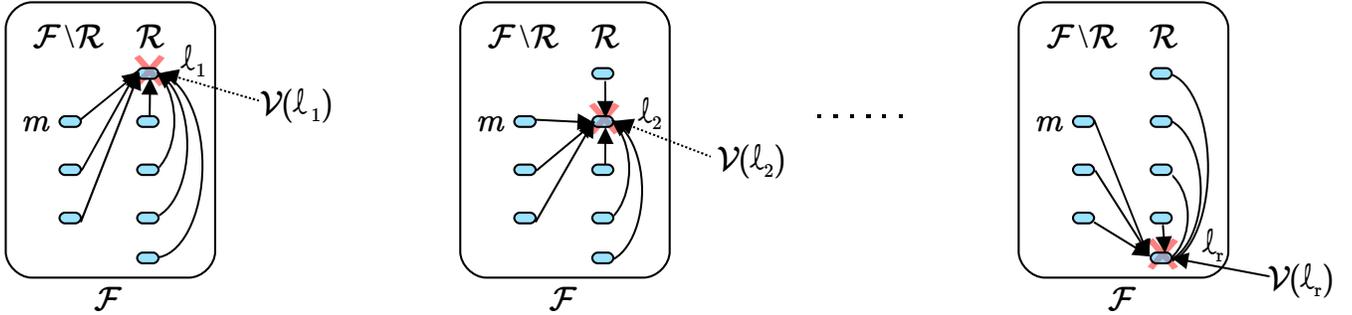

Fig. 5: Helper node pooling, and the setting of Properties 4 and 5.

Regenerating codes must necessarily satisfy helper node pooling scenarios. This leads to surprising (and as we shall see, implausible) upper bounds on the amount of information passed by a single helper node in the pool to multiple replacement nodes. In the following two properties, $S_m^\ell$ is used to denote the random variable corresponding to the data passed by node $m \in \mathcal{F}\backslash\mathcal{R}$ to assist in the repair of node $\ell \in \mathcal{R}$ in the scenario where the $d$ helper nodes are $\{(\mathcal{F}\backslash\{\ell\}) \cup \mathcal{V}(\ell)\}$ [7] . Further, in agreement with our earlier notation, we define

$$S_m^\mathcal{R} \triangleq \left\{ S_m^\ell \right\}_{\ell \in \mathcal{R}} .$$

*Property 4:* In the helper node pooling scenario where

$$\min\{k, f\} > p + 2 \geq r,$$

for an arbitrary node $m \in \mathcal{F}\backslash\mathcal{R}$ it must be that

$$H\left(S_m^\mathcal{R}\right) \leq 2\beta - \theta. \tag{42}$$

*Property 5:* In the helper node pooling scenario where

$$\min\{k, f\} > p + 1 \geq r \geq 2,$$

for an arbitrary node $m \in \mathcal{F}\backslash\mathcal{R}$ and an arbitrary pair of nodes $\{\ell_1, \ell_2\} \in \mathcal{R}$, it must be that

$$H(S_m^{\ell_1} \mid S_m^{\ell_2}) \leq \theta \tag{43}$$

---

[6] This notation is in anticipation of the usage of these sets in Properties 4 and 5, which consider the repair of each of the nodes in $\mathcal{R}$. In the scenario considered, every *replacement* node in $\mathcal{R}$ connects to the remaining $(f-1)$ nodes in $\mathcal{F}$, and hence $\mathcal{F}$ forms a *fixed* set of helper nodes. On the other hand, the set $\mathcal{V}(\ell)$ comprising of the $(d-(f-1))$ helper nodes of node $\ell \in \mathcal{R}$ is specific to node $\ell$ and is allowed to *vary* with $\ell$.

[7] The set $\mathcal{V}(\ell)$ representing the $(d-(f-1))$ arbitrary helper nodes assisting in the repair of node $\ell$ plays no role in the properties or the proof. Hence, for ease of understanding, the reader may choose to assume this set also to be fixed, i.e., $\mathcal{V}(\ell) = \mathcal{V}$.

and hence

$$H\left(S_m^{\mathcal{R}}\right) \leq \beta + (r-1)\theta. \tag{44}$$

Note that Properties 4 and 5 do not hold for the MSR point (which has $p = k - 1$) and are trivially satisfied for the MBR point (which has $p = 0$).

### E. The Non-existence Proof

We now show that the properties derived in the preceding subsection over-constrain the system, causing a majority of the points on the storage-bandwidth tradeoff to be non-achievable under exact-repair. Recall that the parameters $(\alpha, \beta)$ at any point on the tradeoff are written as

$$\alpha = (d-p)\beta - \theta,$$

with the interior points corresponding to

$$p \in \{0, \ldots, k-2\} \quad \text{and} \quad \theta \in [0, \beta), \quad \text{except for the point } \{p = 0, \theta = 0\}.$$

We first consider the case when $\alpha$ is a multiple of $\beta$, in Theorem 6. A majority of regenerating schemes and code constructions in the literature [7]–[13], [16], [17] are designed for this case of $\alpha$ being a multiple of $\beta$. Thus for this case, at the interior points it must be that

$$\alpha = (d-p)\beta, \quad \theta = 0$$

with $p$ lying in the range

$$1 \leq p \leq k-2. \tag{45}$$

*Theorem 6:* For any given values of $B$ and $[n, k, d]$, exact-repair codes do not exist for the parameters $(\alpha, \beta)$ lying in the interior of the storage-bandwidth tradeoff when $\theta = 0$.

*Proof:* The proof is by contradiction: given values of $B$, $[n, k, d]$ and $(\alpha, \beta)$ such that $\theta = 0$, we assume the existence of an exact-repair code with these parameters and show that the code fails to satisfy some of the properties of exact-repair codes.

Let $\mathcal{G}$ denote the distributed storage network under consideration, and further let $\mathcal{F}$ denote an arbitrary sub-network of $\mathcal{G}$ consisting of $(d+1)$ or greater nodes [8]. Given an optimal exact-repair code for the network $\mathcal{G}$, it is clear that the code is also an optimal exact-repair code for the sub-network $\mathcal{F}$, with the same parameter values $k$, $d$, $B$, $\alpha$, $\beta$. In the present proof, we restrict our attention to a sub-network $\mathcal{F}$ consisting of precisely $(d+1)$ nodes.

---

[8] As the notation suggests, in this proof, $\mathcal{F}$ plays the role of the fixed set in the helper node pooling scenario.

A brief outline of the proof is as follows. Clearly, the nodes in $\mathcal{F}$ form a helper node pool, and hence Property 5 can be used to upper bound the total amount of information that a node can pass to aid in the repair of a bunch of nodes. This limits the cumulative information received by the nodes in $\mathcal{F}$ during their respective repair operations, which in turn limits the total data $B$ stored in the network to $(d+1)\beta$. Finally, the value of $B$ determined by the storage-bandwidth tradeoff (3) is found to be strictly larger.

Since the sub-network $\mathcal{F}$ consists of $(d+1)$ nodes, a failed node $\ell \in \mathcal{F}$ is repaired with the assistance of the $d$ remaining nodes in $\mathcal{F}$. Thus for any node $\ell \in \mathcal{F}$, the exact-repair property requires

$$H(W_\ell \mid S^\ell_{\mathcal{F}\setminus\{\ell\}}) = 0. \tag{46}$$

Also, for any three distinct nodes $\{m,\,\ell_1,\,\ell_2\} \in \mathcal{F}$, Property 5 when $\theta = 0$ implies that

$$H(S^{\ell_1}_m \mid S^{\ell_2}_m) = 0. \tag{47}$$

Now, for a data-collector to be able to recover the entire data by connecting to a set of $k$ nodes $\mathcal{K} \subset \mathcal{F}$ it must be that

$$
\begin{aligned}
B \;=\;& H(W_\mathcal{K}) & (48) \\
\leq\;& H(W_\mathcal{F}) & (49) \\
=\;& I\left(W_\mathcal{F};\, \left\{S^\ell_{\mathcal{F}\setminus\{\ell\}}\right\}_{\ell \in \mathcal{F}}\right) & (50) \\
\leq\;& H\left(\left\{S^\ell_{\mathcal{F}\setminus\{\ell\}}\right\}_{\ell \in \mathcal{F}}\right) & (51) \\
=\;& H\left(\left\{S^{\mathcal{F}\setminus\{m\}}_m\right\}_{m \in \mathcal{F}}\right) & (52) \\
\leq\;& \sum_{m \in \mathcal{F}} H\left(S^{\mathcal{F}\setminus\{m\}}_m\right) & (53) \\
\leq\;& \sum_{m \in \mathcal{F}} \beta & (54) \\
=\;& (d+1)\beta \;. & (55)
\end{aligned}
$$

Here, equation (50) follows from the exact-repair requirement stated in equation (46). Equation (52) is a re-writing of equation (51), and equation (54) employs equation (47) and Property 3.

On the other hand, since any optimal regenerating code must satisfy the storage-bandwidth tradeoff (3),

it must be that

$$
\begin{align}
B &= \sum_{i=0}^{k-1} \min\left(\alpha, (d-i)\beta\right) \tag{56} \\
&= \sum_{i=0}^{k-1} \min\left((d-p)\beta, (d-i)\beta\right) \tag{57} \\
&= 2(d-p)\beta + \sum_{i=2}^{k-1} \min\left((d-p)\beta, (d-i)\beta\right) \tag{58} \\
&\geq 2(d-p)\beta + (k-2)\beta \tag{59} \\
&\geq (d+2)\beta \ , \tag{60}
\end{align}
$$

where equation (57) holds since $\alpha = (d-p)\beta$, equation (58) holds since $p \geq 1$ (see equation (45)), and equations (59) and (60) are derived using $d \geq k \geq p + 2$. This is in contradiction to equation (55). ■

*Theorem 7:* For any given values of $B$ and $[n,\ k,\ d]$, exact-repair codes do not exist for the parameters $(\alpha,\ \beta)$ lying in the interior of the storage-bandwidth tradeoff when $\theta \neq 0$ except possibly for the case

$$
p = k-2 \text{ with } \left( \text{either } \theta \geq \frac{d-p-1}{d-p}\beta \ \text{ or } \ k=2 \right).
$$

*Proof:* The proof of this theorem also exploits the existence of helper node pools in the system. Please refer to Appendix B for the proof. ■

*Remark 5:* It can be verified that the properties derived in Section IV-D, and the non-achievability results in the present section continue to hold even if optimal exact-repair of only $k$ of the nodes is desired, and the remaining $n - k$ nodes are permitted to repair functionally with no restriction on the repair bandwidth.

## F. An Achievable Curve via Storage Space Sharing

We have seen that for a majority of the points in the interior of the tradeoff curve, the cut-set bound cannot be achieved under exact-repair. On the other hand, from the codes provided in [16], the cut-set bound can be achieved at the extreme points of the tradeoff curve: for all $[n,\ k,\ d]$ at the MBR point and for $[n,\ k,\ d \geq 2k-2]$ at the MSR point. A linear storage-space-sharing scheme between these codes can be used to establish an achievable region in the interior of the tradeoff curve. Given the system parameters $[n,\ k,\ d]$ and $(B,\ \alpha)$, with $d \geq 2k-2$, the net repair bandwidth $d\beta$ required under this scheme can be computed as

$$
d\beta = \frac{2B - k\alpha}{k(d-k+1)}. \tag{61}
$$

Fig. 6 depicts the curve achieved via storage-space-sharing alongside the storage-bandwidth tradeoff curve for the parameters $[n > 18,\ k = 10,\ d = 18,\ B = 27000]$.

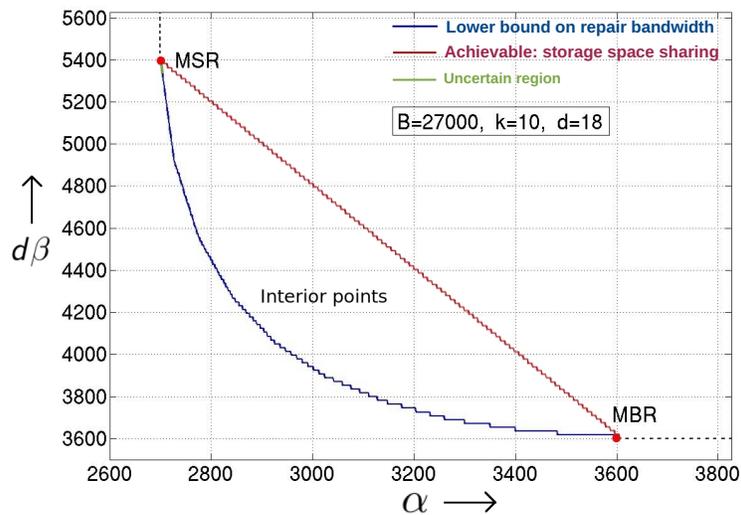

Fig. 6: An achievable value of repair bandwidth $d\beta$ for exact-repair of all nodes plotted alongside the storage repair-bandwidth tradeoff curve, which is a lower bound on the repair bandwidth.

## V. Conclusion

In this paper, an explicit exact MBR code for the parameters $[n, \ k, \ d = n - 1]$ is presented. This code has very low repair complexity; repair of a failed node can be achieved by mere transfer of data and does not require any computation. Moreover, this code, when specialized to the parameter set $[n, \ k = n - 2, \ d = n - 1]$, can be constructed over the binary field, and can be implemented using solely XOR operations.

A set of properties that any exact-repair code must necessarily satisfy are derived. Specific scenarios termed *helper node pooling* are identified, which lead to upper bounds (that are surprisingly small) on the amount of information that a node can pass to assist in the repair of a bunch of nodes. These upper bounds are then used to show the non-achievability of almost all interior points on the storage-bandwidth tradeoff under exact-repair.

## References

[1] D. A. Patterson, G. Gibson, and R. H. Katz, "A Case for Redundant Arrays of Inexpensive Disks (RAID)," in *Proc. ACM SIGMOD*, Chicago, USA, June 1988.

[2] S. Rhea, P. Eaton, D. Geels, H. Weatherspoon, B. Zhao, and J. Kubiatowicz, "Pond:the OceanStore Prototype," in *Proc. USENIXFile and Storage Technologies (FAST)*, 2003.

[3] R. Bhagwan, K. Tati, Yu Chung Cheng, S. Savage, and G. M. Voelker, "Total Recall: System Support for Automated Availability Management," in *NSDI*, 2004.

[4] A. G. Dimakis, P. B. Godfrey, M. Wainwright, and K. Ramchandran, "Network Coding for Distributed Storage Systems," *Proc. IEEE INFOCOM*, Anchorage, May 2007.

[5] Y. Wu, A. G. Dimakis, and K. Ramchandran, "Deterministic Regenerating Codes for Distributed Storage," in *Proc. Allerton Conf.*, Urbana-Champaign, Sep. 2007.

[6] Y. Wu, "Existence and Construction of Capacity-Achieving Network Codes for Distributed Storage," in *Proc. IEEE ISIT,* Seoul, Jul. 2009.

[7] Y. Wu and A. G. Dimakis, "Reducing Repair Traffic for Erasure Coding-Based Storage via Interference Alignment," in *Proc. IEEE ISIT,* Seoul, Jul. 2009.

[8] K. V. Rashmi, N. B. Shah, P. V. Kumar, and K. Ramchandran, "Explicit Construction of Optimal Exact Regenerating Codes for Distributed Storage," in *Proc. Allerton Conf.*, Urbana-Champaign, Sep. 2009.

[9] D. Cullina, A. G. Dimakis and T. Ho, "Searching for Minimum Storage Regenerating Codes," in *Proc. Allerton Conf.*, Urbana-Champaign, Sep. 2009.

[10] N. B. Shah, K. V. Rashmi, P. V. Kumar, and K. Ramchandran, "Explicit Codes Minimizing Repair Bandwidth for Distributed Storage," in *Proc. IEEE ITW,* Cairo, Jan. 2010.

[11] N. B. Shah, K. V. Rashmi, P. V. Kumar, and K. Ramchandran, "Interference Alignment in Regenerating Codes for Distributed Storage: Necessity and Code Constructions," submitted to *IEEE Transactions on Information Theory*, available online at `arXiv:1005.1634v3 [cs.IT]`.

[12] C. Suh and K. Ramchandran, "Interference Alignment Based Exact Regeneration Codes for Distributed Storage," in *Proc. IEEE ISIT,* Austin, Jun. 2010.

[13] Y. Wu, "A Construction of Systematic MDS Codes with Minimum Repair Bandwidth," submitted to *IEEE Transactions on Information Theory*, available online at `arXiv:0910.2486v1 [cs.IT]`.

[14] V. R. Cadambe, S. A. Jafar, and H. Maleki, "Distributed Data Storage with Minimum Storage Regenerating Codes - Exact and Functional Repair are Asymptotically Equally Efficient," available online at `arXiv:1004.4299v1 [cs.IT]`.

[15] C. Suh and K. Ramchandran, "On the Existence of Optimal Exact-Repair MDS Codes for Distributed Storage," available online at `arXiv:1004.4663v1 [cs.IT]`.

[16] K. V. Rashmi, N. B. Shah, and P. V. Kumar, "Optimal Exact-Regenerating Codes for the MSR and MBR Points via a Product-Matrix Construction," submitted to *IEEE Transactions on Information Theory*, available online at `arXiv:1005.4178 [cs.IT]`.

[17] B. Gaston and J. Pujol, "Double Circulant Minimum Storage Regenerating Codes," available online at `arXiv:1007.2401 [cs.IT]`.

[18] F Oggier and A Datta, "Self-repairing Homomorphic Codes for Distributed Storage Systems," available online at `arXiv:1008.0064 [cs.IT]`.

[19] N. B. Shah, K. V. Rashmi, and P. V. Kumar "A Flexible Class of Regenerating Codes for Distributed Storage," in *Proc. IEEE ISIT,* Austin, Jun. 2010.

[20] A.-M. Kermarrec, N. Le Scouarnec and G. Straub, "Beyond Regenerating Codes," Research Report, Institut National De Recherche En Informatique Et En Automatique.

[21] Y. Hu, Y. Xu, X. Wang, C. Zhan and P. Li, "Cooperative Recovery of Distributed Storage Systems from Multiple Losses with Network Coding," *IEEE Journal on Selected Areas in Communication*, vol. 28, no. 2, pp.268276, Feb, 2010.

[22] S. Akhlaghi, A. Kiani and M. R. Ghanavati, "A Fundamental Trade-off Between The Download Cost And Repair Bandwidth In Distributed Storage Systems," in *Proc. IEEE International Symposium on Network Coding (NetCod)*, Toronto, Jun. 2010.

[23] S. El Rouayheb and K. Ramchandran, "Fractional Repetition Codes for Repair in Distributed Storage Systems," in *Proc. Allerton Conf.*, Urbana-Champaign, Sep. 2010.

[24] http://www.ece.iisc.ernet.in/~vijay/storage

# Appendix A
## Proofs of the Properties of Exact-Repair Codes

*Proof of Property 1:* Without loss of generality, assume $\ell = 1$. Now, for reconstruction by a data-collector connecting to the first $k$ nodes, it must be that

$$B = H\left(W_{[k]}\right) \tag{62}$$

$$= H(W_1) + \sum_{j=1}^{k-1} H\left(W_{j+1} \,\middle|\, W_{[j]}\right) \tag{63}$$

$$\leq \alpha + \sum_{j=1}^{k-1} H\left(W_{j+1} \,\middle|\, W_{[j]}\right) \tag{64}$$

$$\leq \alpha + \sum_{j=1}^{k-1} \min\{\alpha, (d-j)\beta\} \tag{65}$$

$$= \sum_{j=0}^{k-1} \min\{\alpha, (d-j)\beta\} \tag{66}$$

$$= B. \tag{67}$$

Here, equation (65) results from Lemma 3, equation (66) uses the fact that $\alpha \leq d\beta$ (equation (36)), and equation (67) follows since we need to satisfy the storage-bandwidth tradeoff with equality (3). Thus, equation (64) must be satisfied with equality, which forces $H(W_1) = \alpha$. ∎

*Proof of Property 2:* The result clearly holds when $a \geq k$ since (i) data contained in any $k$ nodes suffice to recover the entire data, and (ii) $H(W_\ell) = \alpha$ (from Property 1).

Now for the case when $a < k$, without loss of generality, we assume that the set $\mathcal{A}$ comprises of the first $a$ nodes in the system and node $\ell$ is the $(a+1)^{th}$ node, i.e., $\mathcal{A} = [a]$ and $\ell = a + 1$. For reconstruction by a data-collector connecting to the first $k$ nodes, we need

$$B = H\left(W_{[k]}\right) \tag{68}$$

$$= \sum_{j=0}^{k-1} H\left(W_{j+1} \,\middle|\, W_{[j]}\right) \tag{69}$$

$$\leq \sum_{j=0}^{k-1} \min\{\alpha, (d-j)\beta\} \tag{70}$$

$$= B, \tag{71}$$

where equation (70) follows from Lemma 3 and and equation (71) is a result of satisfying the storage-bandwidth tradeoff with equality (3). Thus, equation (70) must be satisfied with equality. This, coupled with the lower bound on each term $H(W_{j+1}|W_{[j]})$ from Lemma 3, gives (for the choice of $j$ as $a$)

$$H\left(W_{a+1}\,\middle|\,W_{[a]}\right) = \min\,(\alpha, (d-a)\beta).\tag{72}$$

Hence,

$$
\begin{aligned}
I\left(W_{a+1}; W_{[a]}\right) &= H(W_{a+1}) - H(W_{a+1}|W_{[a]}) &(73)\\
&= \alpha - \min\,(\alpha, (d-a)\beta) &(74)\\
&= (\alpha - (d-a)\beta)^+ &(75)\\
&= ((a-p)\beta - \theta)^+\,, &(76)
\end{aligned}
$$

where equation (74) follows from Property 1 and equation (72). ∎

*Proof of Corollary 5:* From Lemma 3, it must be that

$$H(W_\ell|S_{\mathcal{A}}^\ell) \le \min\{\alpha, (d-a)\beta\}.\tag{77}$$

On the other hand, since $H(S_m^\ell|W_m) = 0$ for all node $m \in \mathcal{A}$,

$$
\begin{aligned}
H(W_\ell|S_{\mathcal{A}}^\ell) &\ge H(W_\ell|W_{\mathcal{A}}) &(78)\\
&= H(W_\ell) - I(W_\ell; W_{\mathcal{A}}) &(79)\\
&= \alpha - (\alpha - (d-a)\beta)^+ &(80)\\
&= \min\{\alpha, (d-a)\beta\}, &(81)
\end{aligned}
$$

where equation (80) employs Property 1 and Property 2 with $a < k$. ∎

*Proof of Property 3:* Partition the set of $d$ helper nodes assisting in the repair of node $\ell$ into a set $\mathcal{A}$ consisting of $(k-1)$ nodes and a second set $\mathcal{B}$ consisting of the remaining $(d-k+1)$ nodes, such that node $m \in \mathcal{B}$. Then, Corollary 5 mandates that

$$H(W_\ell|S_{\mathcal{A}}^\ell) = (d-k+1)\beta.\tag{82}$$

However, exact-repair of node $\ell$ requires

$$H(W_\ell|S_{\mathcal{A}}^\ell, S_{\mathcal{B}}^\ell) = 0.\tag{83}$$

Equations (82) and (83) imply that

$$H\left(S_{\mathcal{B}}^\ell\right) \ge (d-k+1)\beta.\tag{84}$$

Noting that each of the $(d-k+1)$ helper nodes considered in equation (84) can pass no more than $\beta$ units of information, it must be that

$$H\left(S_{\mathcal{B}}^\ell\right) = (d-k+1)\beta,\tag{85}$$

from which it follows that

$$H\left(S_m^\ell\right) = \beta. \tag{86}$$

∎

*Proof of Property 4:* Clearly, if the statement holds for some values of $f$ and $r$, it continues to hold for all $f'$, $r'$ when $f' \geq f$ and $r' \leq r$. Hence, throughout the proof, the set $\mathcal{R}$ is assumed to be comprised of $r = p+2$ nodes, and the set $\mathcal{F}$ is such that $\mathcal{F} = \mathcal{R} \cup \{m\}$. Thus, $f = p+3$.

Consider repair of an arbitrary node $\ell \in \mathcal{R}$ where the set of $d$ helper nodes includes node $m$ and the $(p+1)$ remaining nodes in $\mathcal{R}$. As an intermediate step, we wish to to prove $H\left(S_m^\ell \big| W_\mathcal{R}\right) = 0$. For this, consider,

$$
\begin{align}
I\left(S_m^\ell; W_\mathcal{R}\right) &= I\left(S_m^\ell; W_\ell, W_{\mathcal{R}\setminus\{\ell\}}\right) \tag{87} \\
&= I\left(S_m^\ell; W_{\mathcal{R}\setminus\{\ell\}}\right) + I\left(S_m^\ell; W_\ell \big| W_{\mathcal{R}\setminus\{\ell\}}\right) \tag{88} \\
&\geq I\left(S_m^\ell; W_\ell \big| W_{\mathcal{R}\setminus\{\ell\}}\right) \tag{89} \\
&= H\left(W_\ell \big| W_{\mathcal{R}\setminus\{\ell\}}\right) - H\left(W_\ell \big| W_{\mathcal{R}\setminus\{\ell\}}, S_m^\ell\right) \tag{90} \\
&\geq H\left(W_\ell \big| W_{\mathcal{R}\setminus\{\ell\}}\right) - H\left(W_\ell \big| S_{\mathcal{R}\setminus\{\ell\}}^\ell, S_m^\ell\right) \tag{91} \\
&= (d-p-1)\beta - (d-p-2)\beta \tag{92} \\
&= \beta, \tag{93}
\end{align}
$$

where equation (91) follows since $S_{\mathcal{R}\setminus\{\ell\}}^\ell$ is a function of $W_{\mathcal{R}\setminus\{\ell\}}$ and equation (92) follows from Property 2 and Corollary 5 with $r = p+2 < k$. Then it must be that,

$$
\begin{align}
H\left(S_m^\ell \big| W_\mathcal{R}\right) &= H\left(S_m^\ell\right) - I\left(S_m^\ell; W_\mathcal{R}\right) \tag{94} \\
&\leq \beta - \beta \tag{95} \\
&= 0, \tag{96}
\end{align}
$$

and hence

$$H\left(S_m^\ell \big| W_\mathcal{R}\right) = 0. \tag{97}$$

Now, since the choice of node $\ell$ from the set $\mathcal{R}$ was arbitrary, equation (97) holds for all $\ell \in \mathcal{R}$ and hence

$$H\left(S_m^\mathcal{R} \big| W_\mathcal{R}\right) = 0. \tag{98}$$

It follows that

$$
\begin{align}
H(S_m^\mathcal{R}) &= I(S_m^\mathcal{R}; W_\mathcal{R}) \tag{99} \\
&\leq I(W_m; W_\mathcal{R}) \tag{100} \\
&= 2\beta - \theta, \tag{101}
\end{align}
$$

where equation (101) follows from Property 2.  ∎

*Proof of Property 5:* The steps followed in this proof are similar to those in the proof of Property 4. Clearly, if the statement holds for some values of $f$ and $r$, it continues to hold for all $f'$, $r'$ when $f' \geq f$ and $r' \leq r$. Hence, throughout the proof, the set $\mathcal{R}$ is assumed to be comprised of $r = p + 1$ nodes, and the set $\mathcal{F}$ is such that $\mathcal{F} = \mathcal{R} \cup \{m\}$. Thus, $f = p + 2$.

Consider repair of an arbitrary node $\ell \in \mathcal{R}$ where the set of $d$ helper nodes includes node $m$ and the $p$ remaining nodes in $\mathcal{R}$. As an intermediate step, we wish to to prove $H\left(S_m^\ell \middle| W_\mathcal{R}\right) \leq \theta$. For this, consider

$$
\begin{align}
I\left(S_m^\ell; W_\mathcal{R}\right) &= I\left(S_m^\ell; W_\ell, W_{\mathcal{R}\setminus\{\ell\}}\right) \tag{102} \\
&= I\left(S_m^\ell; W_{\mathcal{R}\setminus\{\ell\}}\right) + I\left(S_m^\ell; W_\ell \middle| W_{\mathcal{R}\setminus\{\ell\}}\right) \tag{103} \\
&\geq I\left(S_m^\ell; W_\ell \middle| W_{\mathcal{R}\setminus\{\ell\}}\right) \tag{104} \\
&= H\left(W_\ell \middle| W_{\mathcal{R}\setminus\{\ell\}}\right) - H\left(W_\ell \middle| W_{\mathcal{R}\setminus\{\ell\}}, S_m^\ell\right) \tag{105} \\
&\geq H\left(W_\ell \middle| W_{\mathcal{R}\setminus\{\ell\}}\right) - H\left(W_\ell \middle| S_{\mathcal{R}\setminus\{\ell\}}^\ell, S_m^\ell\right) \tag{106} \\
&= (d-p)\beta - \theta - (d-p-1)\beta \tag{107} \\
&= \beta - \theta, \tag{108}
\end{align}
$$

where equation (106) follows since $S_{\mathcal{R}\setminus\{\ell\}}^\ell$ is a function of $W_{\mathcal{R}\setminus\{\ell\}}$ and equation (107) follows from Property 2 and Corollary 5 with $r = p + 1 < k$. Then, it must be that

$$
\begin{align}
H\left(S_m^\ell \middle| W_\mathcal{R}\right) &= H\left(S_m^\ell\right) - I\left(S_m^\ell; W_\mathcal{R}\right) \tag{109} \\
&\leq \beta - (\beta - \theta) \tag{110} \\
&= \theta. \tag{111}
\end{align}
$$

Since the choice of node $\ell$ from the set $\mathcal{R}$ was arbitrary, the above equation holds for all $\ell \in \mathcal{R}$.

Next, we prove $H(S_m^{\ell_1} | S_m^{\ell_2}) \leq \theta$ for an arbitrary pair of nodes $\{\ell_1, \ell_2\} \in \mathcal{R}$. For this, consider

$$
\begin{align}
H(S_m^{\ell_1}, S_m^{\ell_2}) &= I(W_\mathcal{R}; S_m^{\ell_1}, S_m^{\ell_2}) + H(S_m^{\ell_1}, S_m^{\ell_2} | W_\mathcal{R}) \tag{112} \\
&\leq I(W_\mathcal{R}; W_m) + H(S_m^{\ell_1}, S_m^{\ell_2} | W_\mathcal{R}) \tag{113} \\
&= I(W_\mathcal{R}; W_m) + H(S_m^{\ell_1} | W_\mathcal{R}) + H(S_m^{\ell_2} \middle| W_\mathcal{R}, S_m^{\ell_1}) \tag{114} \\
&\leq (\beta - \theta) + \theta + \theta \tag{115} \\
&= \beta + \theta, \tag{116}
\end{align}
$$

where equation (113) follows since $S_m^{\ell_1}$ and $S_m^{\ell_2}$ are functions of $W_m$ and equation (115) follows from

Property 2, and equation (111). Then it must be that,

$$
\begin{aligned}
H(S_m^{\ell_1}|S_m^{\ell_2}) &= H(S_m^{\ell_1}, S_m^{\ell_2}) - H(S_m^{\ell_1}) & (117)\\
&\leq (\beta + \theta) - \beta & (118)\\
&= \theta. & (119)
\end{aligned}
$$

Finally, ordering the nodes in $\mathcal{R}$ in an arbitrary manner as $\{\ell_i \mid 1 \leq i \leq r\}$ and noting that equation (118) holds for every pair of nodes in $\mathcal{R}$, we have

$$
\begin{aligned}
H(S_m^{\mathcal{R}}) &\leq H(S_m^{\ell_1}) + \sum_{i=2}^{r} H(S_m^{\ell_i}|S_m^{\ell_{i-1}}) & (120)\\
&\leq \beta + (r-1)\theta. & (121)
\end{aligned}
$$

∎

## Appendix B
## Proof of Theorem 7

*Proof:* The proof is by contradiction: given values of $B$, $[n, \ k, \ d]$ and $(\alpha, \ \beta)$ such that $\theta \neq 0$, we assume the existence of an exact-repair code with these parameters and show that the properties of exact-repair codes lead to a contradiction.

As in the proof in Theorem 6, we restrict our attention to a sub-network $\mathcal{F}$ of the distributed storage network $\mathcal{G}$ consisting of $(d+1)$ nodes, and ignore the remaining nodes in $\mathcal{G}$. Thus, on failure of a node in this sub-network $\mathcal{F}$, the $d$ helper nodes comprise of the $d$ remaining nodes in $\mathcal{F}$.

Let $\{\ell, m\}$ be a pair of nodes in the sub-network $\mathcal{F}$. Partition the $(d-1)$ remaining nodes in $\mathcal{F}$ into two sets, namely $\mathcal{A}$ of cardinality $p$ and $\mathcal{B}$ of cardinality $(d-p-1)$.

Since the sub-network $\mathcal{F}$ consists of $(d+1)$ nodes, a failed node $\ell \in \mathcal{F}$ is repaired with the assistance of the $d$ remaining nodes in $\mathcal{F}$. Exact-repair of nodes $\ell$ and $m$ requires

$$
\begin{aligned}
H\left(W_\ell \,\middle|\, S_\mathcal{A}^\ell, S_\mathcal{B}^\ell, S_m^\ell\right) &= 0, & (122)\\
H\left(W_m \,\middle|\, S_\mathcal{A}^m, S_\mathcal{B}^m, S_\ell^m\right) &= 0, & (123)
\end{aligned}
$$

respectively. However, since $S_i^m$ is a function of $W_i$ for every helper node $i$, equations (122) and (123) lead to

$$
H\left(W_\ell, W_m \,\middle|\, W_\mathcal{A}, S_\mathcal{B}^\ell, S_\mathcal{B}^m, S_m^\ell\right) = 0. \qquad (124)
$$

Now,

$$
\begin{align}
H(S_{\mathcal{B}}^\ell, S_{\mathcal{B}}^m, S_m^\ell) \;&\geq\; I(S_{\mathcal{B}}^\ell, S_{\mathcal{B}}^m, S_m^\ell; W_\ell, W_m | W_{\mathcal{A}}) \tag{125}\\
&\geq\; H\left(W_\ell, W_m | W_{\mathcal{A}}\right) \tag{126}\\
&=\; H\left(W_\ell | W_{\mathcal{A}}\right) + H\left(W_m | W_{\mathcal{A}}, W_\ell\right) \tag{127}\\
&=\; H(W_\ell) - I\left(W_\ell; W_{\mathcal{A}}\right) + H(W_m) - I\left(W_m; W_{\mathcal{A}}, W_\ell\right) \tag{128}\\
&=\; \alpha - 0 + \alpha - (\beta - \theta) \tag{129}\\
&=\; (2d - 2p - 1)\beta - \theta, \tag{130}
\end{align}
$$

where equation (126) follows from equation (124), and equation (129) follows from Property 1 and Property 2.

Next, we obtain an upper bound on the quantity $H\left(S_{\mathcal{B}}^\ell, S_{\mathcal{B}}^m, S_m^\ell\right)$. We consider the case of $p + 2 < k$ and the case of $p + 2 = k$ separately.

*Case 1:* $p + 2 < k$

In this case,

$$
\begin{align}
H\left(S_{\mathcal{B}}^\ell, S_{\mathcal{B}}^m, S_m^\ell\right) \;&\leq\; \sum_{i \in \mathcal{B}} H\left(S_i^\ell, S_i^m\right) + H\left(S_m^\ell\right) \tag{131}\\
&\leq\; \sum_{i \in \mathcal{B}} (2\beta - \theta) + \beta \tag{132}\\
&=\; (2d - 2p - 1)\beta - (d - p - 1)\theta. \tag{133}
\end{align}
$$

where equation (132) follows from Properties 3 and 4. Since $\theta \neq 0$ and $d \geq k > p + 2$, equations (130) and (133) are in contradiction.

*Case 2:* $p + 2 = k$

In this case, Property 5 is used to obtain an upper bound on $H\left(S_{\mathcal{B}}^\ell, S_{\mathcal{B}}^m, S_m^\ell\right)$. Note that this property does not hold when $k = 2$, and hence we consider the case when $k > 2$.

$$
\begin{align}
H\left(S_{\mathcal{B}}^\ell, S_{\mathcal{B}}^m, S_m^\ell\right) \;&\leq\; \sum_{i \in \mathcal{B}} H\left(S_i^\ell, S_i^m\right) + H\left(S_m^\ell\right) \tag{134}\\
&\leq\; \sum_{i \in \mathcal{B}} (\beta + \theta) + \beta \tag{135}\\
&=\; (d - p)\beta + (d - p - 1)\theta. \tag{136}
\end{align}
$$

where equation (135) follows from Properties 3 and 5. Clearly, equations (136) and (130) contradict when

$$
\theta < \frac{d - p - 1}{d - p}\beta. \tag{137}
$$

∎